\documentclass[twocolumn,english,aps,prl,amsmath,amssymb,reprint,floatfix,showkeys]{revtex4-2}

\usepackage[table]{xcolor}
\usepackage{graphicx}
\usepackage{dcolumn}
\usepackage{bm}
\usepackage[colorlinks = true,citecolor=blue,linkcolor = blue,urlcolor=blue]{hyperref}

\usepackage[noabbrev]{cleveref}
\usepackage{graphicx}	
\usepackage{amsmath}	
\usepackage{mathtools}
\usepackage{amssymb}	
\usepackage{wasysym}    
\usepackage{mathtools}  
\usepackage{aas_macros}
\usepackage{array,multirow,graphicx}  

\newcommand{\Hunit}{km\,s$^{-1}$\,Mpc$^{-1}$}

\begin{document}

\preprint{XXX}

\title{A new method to determine $H_0$ from cosmological energy-density measurements}

\author{Alex Krolewski}
 \email{alex.krolewski@uwaterloo.ca}
 \author{Will J. Percival}%
  \affiliation{Waterloo Centre for Astrophysics, University of Waterloo, Waterloo, ON N2L 3G1, Canada}
  \affiliation{Department of Physics and Astronomy, University of Waterloo, Waterloo, ON N2L 3G1, Canada}
  \affiliation{Perimeter Institute for Theoretical Physics, 31 Caroline St. North, Waterloo, ON NL2 2Y5, Canada}
 \author{Alex Woodfinden}%
  \affiliation{Waterloo Centre for Astrophysics, University of Waterloo, Waterloo, ON N2L 3G1, Canada}
  \affiliation{Department of Physics and Astronomy, University of Waterloo, Waterloo, ON N2L 3G1, Canada}

\date{\today}

\begin{abstract}
We introduce a new method for measuring the Hubble parameter from low-redshift large-scale observations that is independent of the comoving sound horizon. The method uses the baryon-to-photon ratio determined by the primordial deuterium abundance, together with Big Bang Nucleosynthesis (BBN) calculations and the present-day CMB temperature, to determine the physical baryon density $\Omega_b h^2$. The baryon fraction $\Omega_b/\Omega_m$ is measured using the relative amplitude of the baryonic signature in galaxy clustering measured by the Baryon Oscillation Spectroscopic Survey, scaling the physical baryon density to the physical matter density. The physical density $\Omega_mh^2$ is then compared with the geometrical density $\Omega_m$ from Alcock-Paczynski measurements from Baryon Acoustic Oscillations (BAO) and voids, to give $H_0$. Including type Ia supernovae and uncalibrated BAO, we measure $H_0 = 67.1^{+6.3}_{-5.3}$ \,\Hunit. We find similar results when varying analysis choices, such as measuring the baryon signature from the reconstructed correlation function, or excluding supernovae or voids. This measurement is currently consistent with both the distance-ladder and CMB $H_0$ determinations, but near-future large-scale structure surveys will obtain 3--4$\times$ tighter constraints.
\end{abstract}

\maketitle


\section{Introduction}
\label{sec:intro}

The standard model in cosmology, flat $\Lambda$CDM, has been the benchmark for the last 25 years. 
Recently, however, tensions have started to appear between the parameters of the model as measured from different data. The strongest of these tensions is that the $H_0$ measurements from Cepheid variables \cite{Riess-Hubble} disagree with those from the CMB \cite{Planck:2020} or Baryon Acoustic Oscillation (BAO) data \cite{Alam-eBOSS:2021,DESI2024.VI.KP7A} analysed together with Big Bang nucleosynthesis (BBN) measurements. It is imperative to understand tensions such as these in case they require extensions to the standard flat $\Lambda$CDM model. 

The Hubble constant is an absolute quantity with units of inverse time and therefore needs to be constrained by an absolute measurement, which for distance ladder measurements typically comes from geometrical measurements of the distance to the LMC (e.g. \cite{LMC-binary}). For the CMB or BAO measurements, this comes from the CMB temperature. Relative measurements of Cepheid brightness as a function of redshift, or of angular anisotropies in the CMB, then allow us to convert these absolute measurements to a measurement of $H_0$. The CMB temperature is extremely well-measured, $T_\gamma = 2.7255 \pm 0.00057$ \cite{Fixsen2009}, and the shift required to solve the Hubble tension is too large to be viable (e.g. \cite{Ivanov-CMB-temp})

The most accurate way of measuring $H_0$ from CMB and large scale structure is by using the sound horizon as a standard ruler.
The comoving sound horizon at the baryon drag epoch, $r_d$, is extremely well-determined
in the standard cosmological model, with the absolute baseline set by the CMB temperature.
But unlike the distance-ladder measurements, this requires an assumption on the cosmological model.
Therefore, a promising way to raise $H_0$ from CMB and LSS is by modifying the cosmological model around recombination, decreasing the sound horizon and increasing $H_0$ \cite{KnoxMillea,Aylor19,Bernal16}.
One model that accomplishes this is Early Dark Energy (EDE), which posits the existence of an additional scalar field in the early Universe, leading to an increase in the expansion rate around the baryon drag epoch \citep{Karwal:2016,Polin:2019}.

We present a new method to measure $H_0$ comparing low redshift physical and geometrical matter densities without having to model the sound horizon. This combines $\Omega_b h^2$ from the CMB temperature and light element abundances together with the baryon fraction $\Omega_b/\Omega_m$ from the amplitude of the baryon signal in the power spectrum. 
Finally, adding a geometrical measurement of the matter density $\Omega_m$ from the Alcock-Paczynski (AP) effect in BAO and voids, uncalibrated standard rulers, and uncalibrated standard candles is sufficient to form a new measurement of $H_0$. 
In this letter, we present the cosmological results of this combination, while our companion paper \citep{krolewski} provides details on the robustness of the baryon fraction measurement from the galaxy power spectrum and performs an extensive series of validation tests on both $N$-body mocks and noiseless theory vectors.

There have been previous attempts to measure $H_0$ by marginalising over $r_s$ to remove the dependence on the sound horizon \cite{BaxterSherwin21,Philcox21c,Philcox22,Farren22,Madhavacheril24}.
These $H_0$ measurements use the matter power spectrum turnover, which occurs at the matter-radiation equality scale. Since EDE alters the shape of the power spectrum, these measurements are potentially unable to rule out EDE despite the sound horizon freedom \cite{Smith22,Kable24}). We provide a new method that extracts the amplitude of the baryon signal, sacrificing signal-to-noise on $H_0$ for simplicity of the underlying physics.
Like the equality-scale measurement, this method is not independent of $H_0$ constraints in full-shape fits to galaxy clustering data \cite{Ivanov19,Colas19}, but rather provides an additional robustness test of $H_0$ constraints.


The new method is discussed in more detail in the next section, and then we consider results from the Sloan Digital Sky Survey \cite{York00} Baryon Oscillation Spectroscopic Survey (BOSS) DR12 sample \cite{Dawson13}. We conclude with a discussion and a look forward to the improved large-scale structure measurements due from the Dark Energy Spectroscopic Instrument (DESI \cite{DESI2016}) and Euclid satellite \cite{Euclid2011}. 

\section{The Hubble constant from galaxy surveys using energy densities}

Our new method to measure the Hubble parameter can be understood as follows. We start with (i) the CMB temperature, which provides a strong (absolute) constraint on the present-day physical photon density $\epsilon_{\gamma,0}\propto\Omega_{\gamma,0}h^2$. Next, observations of primordial deuterium abundances together with BBN calculations tell us (ii) the photon-to-baryon ratio. The amplitude of the baryonic signal in the galaxy power spectrum constrains (iii) the ratio between the baryon density and the matter density.
Distance measurements from BAO and voids constrain (iv) the matter density relative to the critical density  $\Omega_{m,0}\equiv\epsilon_{m,0}/\epsilon_{c,0}$ 
from geometrical considerations at low redshift.
The covariance between this measurement and the baryon fraction is included in the analysis and contributes negligibly to our $H_0$ errorbar. 
Our method is explicitly independent of the sound horizon, allowing it to test Hubble-tension resolving models such as Early Dark Energy as well as modified recombination histories \cite[e.g.][]{SchoenebergVacher24,Mirpoorian24}.

The four measurements are used to measure $H_0$ as schematically summarised in Eq.~\ref{eq:method}
\begin{equation}  \label{eq:method}
    \frac{3c^2H_0^2}{8\pi G} = \epsilon_c=
    \underbrace{\vphantom{
\frac{\epsilon_{b,0}}{\epsilon_{\gamma,0}} 
    }
    \epsilon_{\gamma,0} }_{\textrm{(i)}}
    \times
    \underbrace{
    \frac{\epsilon_{b,0}}{\epsilon_{\gamma,0}}}_{\textrm{(ii)}}
    \times
    \underbrace{\frac{\epsilon_{m,0}}{\epsilon_{b,0}}}_{\textrm{(iii)}}
    \times
    \underbrace{\frac{1}{\Omega_{m,0}}}_{\textrm{(iv)}}\,, 
\end{equation}
with each portion numbered according to the paragraph above
and $\epsilon$ indicating physical energy densities.

The method relies on having a Universe in which the CMB is a blackbody, 
BBN is described by standard-model physics,
and structure in the matter and galaxy fields depends on the baryon and CDM components in proportion to their relative densities. We also assume that the BAO and voids are isotropic, the BAO (SNe) are standard rulers (standardizable candles), the Universe is flat, and dark energy is a cosmological constant, to constrain $\Omega_m$ from the AP measurements.
We consider the amount of curvature and dynamical dark energy required to affect this measurement later. 
The BAO amplitude constrains the ratio of the baryon and CDM densities
$f_b \equiv 
\Omega_b/\Omega_{bc}$ (where $\Omega_{bc}$ is the sum of the baryon and CDM densities),
since galaxy clustering responds to the power spectrum of matter perturbations excluding neutrinos \cite{Villaescusa-Navarro14}. On the other hand, geometric measurements of $\Omega_m$ are also sensitive to the small density of non-relativistic massive neutrinos $\Omega_\nu$; hence our $H_0$ measurement requires an assumption about the neutrino mass. However, the variation in $\sum m_\nu$ required to affect $H_0$ is large compared to current uncertainties \cite{Planck:2020}.

The new part of our $H_0$ constraint is the baryon fraction measurement from galaxy clustering data.
The relative density of baryons in the galaxy clustering signal is usually marginalised in any BAO measurement, as the cosmological constraints it provides are weak compared with those from the CMB. However, the amplitude can be separated from other parameters and depends on the properties of the Universe independent of $r_d$.
Specifically, we measure the baryon fraction by inserting a new parameter that re-weights the baryon and CDM contributions to the transfer function \cite{HuSugiyama96,Eisenstein-Hu} in the linear power spectrum.

\section{Measurements}

We extract a marginalised likelihood for the baryon fraction (part (iii) of Eq.~\ref{eq:method}) from fits to BOSS DR12 galaxy clustering in the companion paper \cite{krolewski}.
That work considers two different methods to estimate $f_b$: template-based models typically used to make BAO measurements \cite{SBRS} fit to 
$\xi(r)$ both before and after BAO reconstruction \cite{Eisenstein07}; and the effective field theory of large scale structure models of \cite{Chudaykin20,Ivanov19} fit to $P(k)$ without reconstruction. 
Based on fits to $N$-body simulations and noiseless theory vectors in EDE cosmologies,  \cite{krolewski} places a systematic error on $f_b$ of 0.013. 
While we present the impact on $H_0$ of using the $f_b$ measurements from both full-shape fits to $P(k)$ and pre- and post-reconstruction fits to $\xi(r)$,
we use the full-shape fits as our baseline  since the EFT model is more flexible and applicable in a wider range of situations. 

To measure $\Omega_m$ (part (iv) of Eq.~\ref{eq:method}), we fit a flat $\Lambda$CDM model to the distance measurements released by the BOSS DR12 team 
\cite{Ross2017,Alam17,Beutler16BAO}. We use their best-fit combined AP (i.e.\ $\alpha_{\parallel}$ and $\alpha_{\perp}$) measurements (Eqs.~21--23 in \cite{Alam17}).
In flat $\Lambda$CDM, these data constrain $\Omega_m$ and $H_0 r_d$; to remain independent of the sound horizon, we only use the $\Omega_m$ constraint, thus yielding an uncalibrated BAO measurement.
 A wide redshift range significantly increases the constraining power on $\Omega_m$ \cite{Lin21,Blomqvist19,Cuceu19,Brieden22}, and hence we also use isotropic BAO measurements from the SDSS Main Galaxy Survey (MGS) at $z = 0.15$ \cite{Ross15}; $D_M$ and $D_H$ from eBOSS quasars at $z=1.48$ \cite{Alam-eBOSS:2021,Hou20,Neveux20}; and $D_M$ and $D_H$ at $z = 2.334$ from the eBOSS Ly$\alpha$ forest autocorrelation and cross-correlation with quasars \cite{duMasdesBourboux20}.

We supplement the $\Omega_m$ constraints from uncalibrated BAO with the void isotropy measurements presented in \cite{Woodfinden2022} for MGS at $0.07 < z < 0.2$ and BOSS DR12 at $0.2 < z < 0.6$. We do not use their highest-redshift measurement, which combines BOSS DR12 and eBOSS at $0.6 < z < 1.0$, because we cannot measure the covariance between this measurement and the galaxy power spectrum at $0.5 < z < 0.75$, due to lack of simulations with suitable geometry.

In our baseline analysis, we add supernova constraints using the Pantheon+ flat $\Lambda$CDM measurement of $\Omega_m = 0.334 \pm 0.018$ as an external prior.
When we consider the impact of a time-varying dark energy or curvature, we instead
represent the 
supernova likelihood
as a 2D or 3D gridded prior on ($\Omega_m$, $\Omega_k$) or ($\Omega_m$, $w_0$, $w_a$) \cite{Brout22}, constructing this from the publicly available Pantheon+ chains \footnote{In detail, we measure the $\Omega_m$-$\Omega_k$ or $\Omega_m$-$w_0$-$w_a$ priors from the ``Pantheon+\_SH0ES'' chains, marginalizing over $H_0$.}. More recent measurements from Union3 ($\Omega_m = 0.36 \pm 0.03$) \cite{UNION} or DES ($\Omega_m = 0.352 \pm 0.017$) \citep{DESSNe} are modestly different from Pantheon+. However, replacing the Pantheon+ prior with a Union3 or DES prior does not appreciably change $H_0$ (Table~\ref{tab:results}).

To estimate (i) and (ii) in Eq.~\ref{eq:method}, we use the recent updated BBN recommendations of \cite{Schoneberg24}, $\Omega_b h^2 = 0.02218 \pm 0.00055$.
This relies on deuterium and helium measurements as compiled by the Particle Data Group (PDG) \cite{PDG} and nuclear rates from the \texttt{PRyMordial} code \cite{Burns24}, which adds marginalization over nuclear reaction rate uncertainties, comfortably encompassing both experimental and ab-initio results.
The observations of deuterium abundance \citep{Cooke2014,Cooke2016,Balashev16,RiemerSorensen17,Cooke18,Zavarygin18,RiemerSorensen15} and helium abundance \cite{Aver21,Valerdi19,Fernandez19,Kurichin21,Hsyu20,Valerdi21} are challenging because the inferred abundances must be primordial, i.e.\ not contaminated by stellar production or destruction of helium and deuterium.

Extensions to the standard model could potentially affect the BBN constraints on $\Omega_b h^2$, although it is insensitive to additional relativistic species $\Delta N_{\textrm{eff}}$ \cite{Schoneberg24}.
The strong consistency of BBN and CMB measurements of $\Omega_b h^2$ ($0.02242 \pm 0.00014$ from Planck18 \cite{Planck:2020}) disfavors models with extra entropy production that modify the photon number density between BBN and recombination \cite[e.g.][]{Simha08}. 
Even in the worst case scenario,
the additional uncertainty in $\Omega_b h^2$ from extended models could only contribute an extra error of $\sim$1 km s$^{-1}$ Mpc$^{-1}$, which is completely negligible for our results which are currently limited by the measurement of the baryon fraction.

The measurements (iii) and (iv) overlap in redshift and thus have a nonzero covariance.
We measure the covariance using 1000 Multi-Dark Patchy mocks \cite{Kitaura16}. For the template-based fits, we measure $f_b$ using our code for all 1000 pre or post-reconstruction mock correlation functions, and measure the covariance with the BOSS team's 1000 AP fits combining post-reconstruction correlation function and power spectrum. We then re-scale the resulting covariance to match the measured error on $f_b$ on data (which is slightly larger than the spread of the mocks). We also measure the covariance with the void measurements by running the void-finding pipeline of \cite{Woodfinden22} on the mocks.

The template-based fits thus consist of two stages, one in which $f_b$, $\alpha_{\parallel}$ and $\alpha_{\perp}$ are measured from the correlation function  \cite{krolewski},
and another in which $H_0$ and $\Omega_m$ are inferred from these measurements.
 This is similar to the two-step process used to infer cosmological parameters from reconstructed BAO \citep[e.g.][]{Ross2017,Alam17}.

\begin{table}[]
    \centering
    \small
    \begin{tabular}{l|cc|c}
    Data Combination & $f_b$ & $\Omega_m$ & $H_0$ \\
    \hline
    \rowcolor{gray!20} Full shape $P(k)$ & $0.153\pm0.026$ & $0.313 \pm 0.012$ & $67.1^{+6.3}_{-5.3}$ \\
    Post-recon. & \multirow{ 2}{*}{$0.170 \pm 0.030$} & \multirow{2}{*}{$0.312 \pm 0.012$} & \multirow{2}{*}{$64.8^{+6.6}_{-5.7}$} \\
    template $\xi(r)$  & & & \\
    Pre-recon. & \multirow{2}{*}{$0.153 \pm 0.032$}
    & \multirow{2}{*}{$0.312 \pm 0.012$} & \multirow{2}{*}{$68.3^{+8.4}_{-6.7}$} \\  
    template $\xi(r)$ & & & \\
    \hline
     Full shape $P(k)$ & \multirow{2}{*}{$0.153\pm0.026$} & \multirow{2}{*}{$0.310 \pm 0.014$} & \multirow{2}{*}{$67.6^{+6.1}_{-5.6}$} \\
     + Union3 SNe & & & \\
     Full shape $P(k)$ & \multirow{2}{*}{$0.153\pm0.026$} & \multirow{2}{*}{$0.323 \pm 0.012$} & \multirow{2}{*}{$66.1^{+6.1}_{-5.4}$} \\
     + DES SNe & & & \\
Full shape $P(k)$ & \multirow{2}{*}{$0.153\pm0.026$} & \multirow{2}{*}{$0.296 \pm 0.015$} & \multirow{2}{*}{$69.1^{+6.1}_{-5.6}$} \\
+ no SNe  & & & \\
      Full Shape $P(k)$  & \multirow{2}{*}{$0.154\pm0.026$} & \multirow{2}{*}{$0.279 \pm 0.015$} & \multirow{2}{*}{$70.9^{+6.6}_{-5.8}$} \\
      + no voids  & & & \\
     Full Shape $P(k)$ & \multirow{2}{*}{$0.150\pm0.026$} & \multirow{2}{*}{$0.385 \pm 0.035$} & \multirow{2}{*}{$61.0^{+6.1}_{-5.5}$} \\
     + BOSS DR12 only & & & \\
     \hline
     Post-recon. & \multirow{2}{*}{$0.175 \pm 0.030$} & \multirow{2}{*}{$0.260^{+0.039}_{-0.054}$} & \multirow{2}{*}{$71.0^{+11.0}_{-7.7}$} \\
     template $\xi(r)$ + $w_0 w_a$ & & & \\
     Post-recon.  &
     \multirow{2}{*}{$0.175 \pm 0.030$} & \multirow{2}{*}{$0.273^{+0.021}_{-0.018}$} & \multirow{2}{*}{$68.4^{+7.3}_{-5.8}$} \\
     template $\xi(r)$  + $\Omega_{\mathrm{k}}$ & & & \\
    \end{tabular}
    \caption{$H_0$ constraints in \Hunit\ from the signature of baryons, uncalibrated BAO positions and supernovae, and BBN. The fiducial $H_0$ constraint is the top (shaded) row. The errors on $f_b$ and $H_0$ include the estimated systematic error of 0.013 in $f_b$ or 2.9 \Hunit in $H_0$, added in quadrature. The next rows show various changes to the baryon amplitude measurement (first three rows, taking $f_b$ fits from \cite{krolewski}), the geometrical datasets (next five rows), or the late-time expansion history (last two rows).
    \label{tab:results}}
\end{table}

In contrast, in the full-shape fits, we simultaneously fit the power spectrum multipoles $P_\ell(k)$ and the correlated distances from the BAO measurements, along with an $\Omega_b h^2$ prior from BBN, to directly obtain $H_0$ constraints in a single step, as is done in EFTofLSS fits \citep[e.g.][]{Ivanov19,DAmico19}.
We therefore use the same fits that were described in \cite{krolewski}, with slight differences as described below.
In this case, we measure the covariance between  $P_\ell(k)$ and the AP parameters from both galaxies and voids. An additional complication is that the power spectra are split between the North Galactic Cap and the South Galactic Cap, 
but the AP parameters are measured on the combined sample. However, all of these quantities can be measured on mocks and the covariance directly computed, similar to \cite{Philcox20}.

In detail, $f_b$ in Table~\ref{tab:results} is slightly different from $f_b$ in Table 6 of \cite{krolewski}. Those fits only use AP measurements from BOSS DR12 BAO, whereas our baseline fits also include supernovae or voids (our ``BOSS DR12 only'' fits include voids as well as BAO to considerably improve the $\Omega_m$ constraints).
Additionally, Table~\ref{tab:results} adds the systematic error on $f_b$ in quadrature while Table 6 of \cite{krolewski} only gives the statistical error.

We use the MCMC samplers \texttt{cobaya} \cite{Cobaya,CobayaCode} for the template fits and \texttt{MontePython} \cite{Brinckmann:2018cvx,Audren:2012wb} for the full-shape fits. Due to dramatically faster likelihood evaluation times for the template fits, we run the template fits to Gelman-Rubin R-1$= 0.01$ \cite{GelmanRubin} and the full-shape fits to Gelman-Rubin R-1$= 0.05$, but we verify in both cases that our results are consistent with stopping at slightly higher values of R-1 = 0.05 or 0.1. We analyze the resulting chains with \texttt{GetDist} \cite{GetDist}.



\section{Results}

\begin{figure}
\includegraphics[scale=0.87]{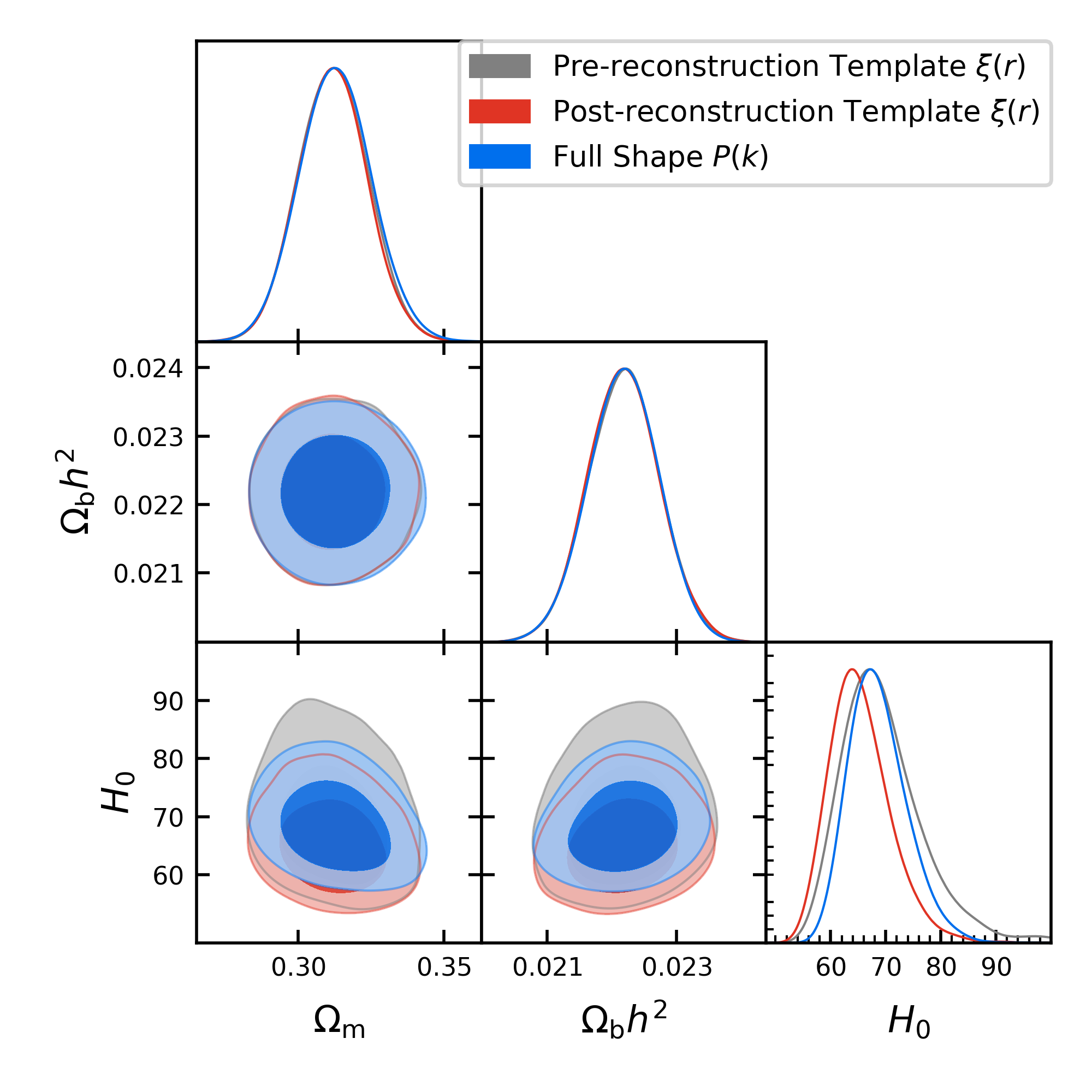}
    \caption{
    Constraints on $H_0$ (in \Hunit), $\Omega_m$ and $\Omega_b h^2$ from the combination of Big Bang Nucleosynthesis; uncalibrated standard rulers, voids, and Type Ia supernovae; and the amplitude of the baryonic signature in BOSS DR12 galaxy clustering. Three variations are shown corresponding to different methods for measuring the baryon fraction: using the EFT of Large Scale Structure on the pre-reconstruction power spectrum (blue); a damped exponential BAO model (with fixed power spectrum template) applied to the pre-reconstruction correlation function (gray); and the same model applied to the post-reconstruction correlation function (red). These contours show statistical errors only, without contribution from systematic errors.
    \label{fig:cosmology_results}}
\end{figure}

In our baseline results, combining the amplitude of the baryonic signature, BBN, uncalibrated standard rulers, and type Ia supernovae, we find $H_0 = 67.1^{+6.3}_{-5.3}$ km s$^{-1}$ Mpc$^{-1}$, including the systematic error budget on $f_b$. The statistical errors alone give $H_0 = 67.1^{+5.6}_{-4.4}$ \Hunit\  with an additional systematic error of $\pm2.9$ \Hunit\  (Fig.~\ref{fig:cosmology_results} and Table~\ref{tab:results}).
The systematic error is derived from scaling the systematic error on $f_b$ to $H_0$ since the $f_b$ errors dominate the errors on $\Omega_m$ or $\Omega_b h^2$.
Our result is robust to changing the method used to measure the baryon fraction: we find a very similar result using the template fit for the pre-reconstruction correlation function, though with a slightly larger error ($68.3^{+8.4}_{-6.7}$\,\Hunit); and a $0.5\sigma$ shift towards lower $H_0$ after reconstruction ($64.8^{+6.6}_{-5.7}$\,\Hunit); in all cases systematic errors are added in quadrature. The difference between the full-shape and post-reconstruction results is consistent with the differences seen in $N$-body mocks \cite{krolewski}.

Indeed, $H_0$ is more sensitive to the geometric dataset used to constrain $\Omega_m$. Both the supernovae and voids prefer high $\Omega_m$, whereas BAO prefers lower $\Omega_m$ (driven by the comparison of BOSS DR12 galaxy and eBOSS Ly$\alpha$ distance measurements). While the supernovae measurements are in $\sim$1$\sigma$ tension with each other, our result does not change much when using Union3 (DES), changing to $67.6^{+6.1}_{-5.6}$\,\Hunit\ ($66.1^{+6.1}_{-5.4}$\,\Hunit). Removing either supernovae or voids lowers $\Omega_m$ and thus raises $H_0$ to $69.1^{+6.1}_{-5.6}$\,\Hunit\ (no SNe) or $70.9^{+6.6}_{-5.8}$\,\Hunit\ (no voids). Finally, when using BOSS DR12 data only (including voids), $\Omega_m$ increases by nearly 2$\sigma$, leading to a very low $H_0$ value of $61.0^{+6.1}_{-5.5}$\,\Hunit. The errors are dominated by the errors on $f_b$ even when the $\Omega_m$ error triples when using BOSS DR12 data alone. 

Finally, this measurement is sensitive to the late-time expansion history. Adding curvature lowers $\Omega_m$ and weakens its constraint, but doesn't change $H_0$ significantly (to $68.4^{+7.3}_{-5.8}$\,\Hunit), a 20\% increase in the error. A varying dark energy equation of state has a larger impact on $H_0$, significantly weakening the constraint in a $w_0$--$w_a$ cosmology to $71.0^{+11.0}_{-7.7}$\,\Hunit. This is largely due to the degeneracy between the poorly constrained $w_a$ and $H_0$.
While we do not explicitly vary the neutrino mass in this work, even using
 the 90\% upper limit on $\Sigma m_\nu$ from direct detection, 0.8 eV \cite{Katrin21}, $\Omega_\nu$ is still a considerably smaller fraction of $\Omega_m$ (5\%) than our $f_b$ fractional error (15\%) and thus will not substantially broaden the $H_0$ constraint.

\section{Discussion}

\begin{figure}
\includegraphics[scale=0.6]{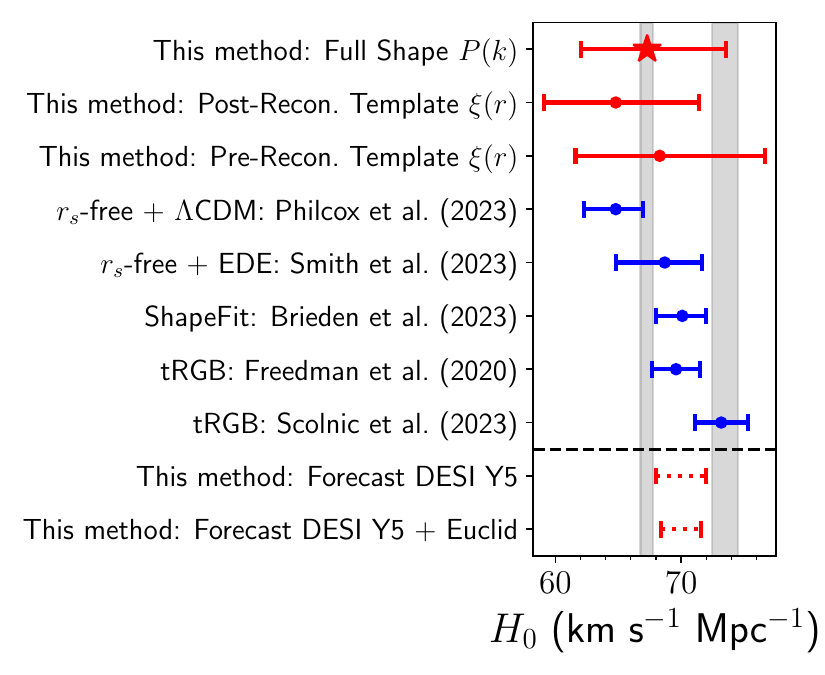}
    \caption{$H_0$ measurements (in \Hunit) from Cepheids \cite{Riess-Hubble} and Planck CMB measurements assuming a flat $\Lambda$CDM cosmology \cite{Planck:2020}, shown in the gray bands. The points show various measurements with minimal additional assumptions: those presented in this paper (red, at top, with fiducial measurement starred); other sound horizon free $H_0$ measurements from large-scale structure; and local $H_0$ measurements from the Tip of the Red Giant Branch. Our constraints on $H_0$ are currently weak and compatible with both the local and CMB/$\Lambda$CDM measurements; but with DESI and Euclid, our method will have sufficient constraining power to differentiate Planck and SH0ES at $\sim$4$\sigma$.
    \label{fig:H0_comparison}}
\end{figure}

Our $H_0$ measurement is consistent both with SH0ES ($73.04\pm1.04$\,\Hunit) \cite{Riess-Hubble} and Planck ($67.26\pm0.49$\,\Hunit) \cite{Rosenberg22} (see also \cite{Tristram24} and \cite{Planck:2020}). In Fig.~\ref{fig:H0_comparison}, we compare to other sound horizon-free $H_0$ measurements from large-scale structure. \cite{Philcox22}, \cite{Smith22} and \cite{Brieden23} infer cosmological parameters from the shape of the matter power spectrum, with explicit marginalization over the sound horizon. While they obtain tighter constraints on $H_0$, these measurements are model-dependent; nonzero Early Dark Energy leads to a different power spectrum shape than in $\Lambda$CDM, shifting the $H_0$ constraint \cite{Smith22,Kable24}. In  contrast, our measurement has larger errors, but is less model-dependent, as demonstrated in our companion paper \cite{krolewski}. For comparison, we also show results from an alternative local calibration of the distance ladder using the Tip of the Red Giant Branch rather than Cepheids. Refs.~\cite{Freedman20,Scolnic23} achieve $H_0$ precision of $\sim$2\,\Hunit, although with a fairly large discrepancy between the two results perhaps suggesting larger systematic errors.

While the current constraining power of this measurement is insufficient to resolve the Hubble tension, ongoing large-scale structure surveys will allow for significantly higher precision. DESI \cite{DESI2016,DESISV} and Euclid \cite{Euclid20} will both survey roughly ten times the effective volume of BOSS DR12. We show in Fig.~\ref{fig:H0_comparison} the forecasted constraining power of these surveys, scaling our fiducial constraint by the square root of the effective volume (as demonstrated in \cite{krolewski}) and centering at 70\,\Hunit.
The systematic errors are dominated by prior effects from the large number of poorly constrained EFT parameters, and hence are expected to improve in tandem with the statistical errors.\footnote{In \cite{krolewski}, we find $\sim 0.5\sigma$ biases on $H_0$ using an EDE cosmology as truth and scaling the covariance by a factor of 10. This is similar to the 0.5$\sigma$ biases that we find with an unscaled covariance, but with a smaller $\sigma_{H_0}$ and thus smaller offset in $H_0$.}
If the systematics remain under control, the combination of DESI and Euclid could reach a precision of 1.6\,\Hunit,\footnote{Removing the $\sim$9000 deg$^2$ overlap between Euclid and DESI from the DESI effective volume, since Euclid has higher number density.} sufficient to differentiate SH0ES and Planck at nearly 4$\sigma$. Moreover, sound horizon free measurements constraining $H_0$ from the turnover of the matter power spectrum will continue to improve \cite{Farren22}. The complementarity of these two large-scale structure measurements could yield intriguing hints about potential solutions to the Hubble tension, or put substantial pressure on currently-allowed models that can resolve the tension by shifting the sound horizon.

\begin{acknowledgments}
We thank Ashley Ross for useful conversations and for helping to dig out old data, and Edmond Chaussidon for useful conversations regarding the DESI/Euclid overlap.
AK was supported as a CITA National Fellow by the Natural Sciences and Engineering Research Council of Canada (NSERC), funding reference \#DIS-2022-568580.
WP acknowledges support from the Natural Sciences and Engineering Research Council of Canada (NSERC), [funding reference number RGPIN-2019-03908] and from the Canadian Space Agency.
Research at Perimeter Institute is supported in part by the Government of Canada through the Department of Innovation, Science and Economic Development Canada and by the Province of Ontario through the Ministry of Colleges and Universities.
This research was enabled in part by support provided by Compute Ontario (computeontario.ca) and the Digital Research Alliance of Canada (alliancecan.ca). 
\end{acknowledgments}

\bibliographystyle{apsrev4-1}
\bibliography{References}


\title{$H_0$ from cosmological energy-density measurements}

\title{H0 from cosmological energy-density measurements}





\date{\today}


\maketitle





\end{document}